\documentclass[twocolumn,english]{revtex4}
\usepackage[T1]{fontenc}
\usepackage[latin1]{inputenc}
\usepackage{graphicx}
\usepackage{amssymb}

\makeatletter

\usepackage{babel}
\makeatother
\begin{document}

\title{Numerical calculation of classical and non-classical electrostatic
potentials}

\author{Dan Christensen, Brian Neyenhuis, Dallin S. Durfee}

\affiliation{Department of Physics and Astronomy, Brigham Young University, Provo,
UT 84602}

\begin{abstract}
We present a numerical exercise in which classical and non-classical
electrostatic potentials were calculated. The non-classical fields
take into account effects due to a possible non-zero photon rest mass.
We show that in the limit of small photon rest mass, both the classical
and non-classical potential can be found by solving Poisson's equation
twice, using the first calculation as a source term in the second
calculation. Our results support the assumptions in a recent proposal
to use ion interferometry to search for a non-zero photon rest mass.
\end{abstract}
\maketitle

\section{Introduction}

Although classical electromagnetism forbids electrostatic fields inside
empty conducting shells, quantum mechanics suggests that small fields
might exist. In the spirit of Yukawa's particle-exchange theory of
forces \cite{Yukawa49Nobel}, a modified version of Maxwell's equations
was derived to account for a possible non-zero rest mass of the photon,
the exchange particle of the Coulomb force \cite{Jackson1975}. Because
a finite photon mass would limit the range of the Coulomb force, these
equations violate Gauss's law and make these fields possible.

The experimental search for deviations from Coulomb's inverse-square
law goes back as early as 1769 \cite{Cavendish1773,Jackson1975,Robinson1769}.
Although no field has been found at the sensitivity level of these
experiments, based on the predicted sensitivity of the experiments
an upper limit on the photon rest mass has been determined. The most
recent tests of Coulomb's law were limited by the sensitivity of the
voltage-measurement electronics and possible back-action of the measurement
process on the potentials being measured \cite{Crandall83,Williams1971}.

Progress on these experiments has been slow --- the limit on the rest
mass of a photon from these types of experiments has decreased by
only a factor of 2.5 in the last 35 years. We recently proposed to
use ion interferometry to improve this measurement by \emph{several
orders of magnitude} \cite{Neyenhuis06cy}. In this experiment, a
voltage would be applied across a concentric pair of conducting cylindrical
shells. A beam of ions traveling through the inner shell would be
split and recombined using either physical gratings or laser beams.
A non-zero electric field in the shell would induce a phase shift
between the arms of the interferometer, resulting in a shift in the
interference pattern.

While investigating the feasibility of the experiment, we performed
several numerical calculations of classical and non-classical electrostatic
potentials in the proposed apparatus. In this paper we discuss the
methods and results of these calculations.

\section{Definition of the Problem}

In the proposed experiment, a non-zero electric field will be searched
for inside the inner of two concentric cylindrical conductors held
at different voltages. The inner conductor's end caps would have several
small holes to allow passage of the ions and to allow laser beams
and/or wires to enter. For the experiment to work, the fringing fields
\char`\"{}leaking\char`\"{} through these holes must be small compared
to the non-classical field under study. In addition, the calculations
in \cite{Neyenhuis06cy} assume that the non-classical part of the
potential between the interferometer gratings is approximately that
of an infinite cylinder. To verify that these conditions are met,
we performed several numerical calculations.

\begin{figure*}
\begin{center}\includegraphics[%
  width=1.5in,
  angle=-90]{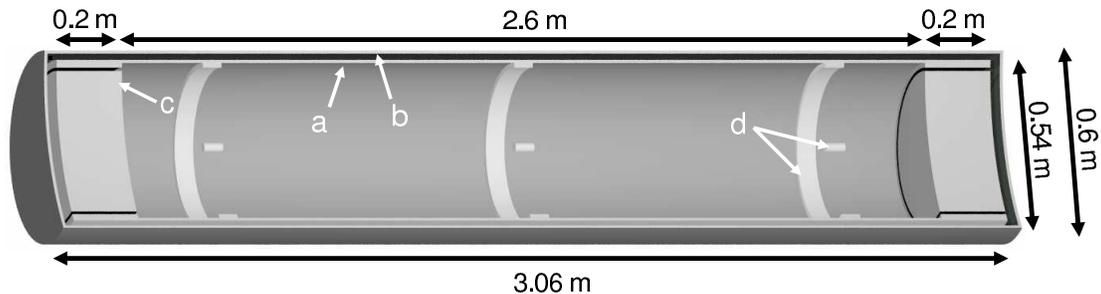}\end{center}

\caption{Layout of conductors. This scale figure gives a cut-away perspective
view of the conductors which enforce the boundary conditions assumed
in our calculations. Arrows point out (a) the inner conducting cylinder,
(b) the outer conducting cylinder, (c) the disk which caps one end
of the inner conductor, and (d) one of the three sets of additional
conductors inserted into our calculation to explore the effect of
mounting objects inside the inner conductor. \label{cap:Layout-of-conductors.}}
\end{figure*}

The setup assumed for these calculations is illustrated in Fig.\
\ref{cap:Layout-of-conductors.}. The inner conducting shell was assumed
to be a thin-walled 2.6 m long cylinder, 27 cm in radius. It is capped
on either end with conducting disks, also 27 cm in radius. These disks
are 20 cm long to reduce fringing fields through the holes. This inner
shell is surrounded by a second conducting tube with end caps. The
outer tube is 3.06 m long with an inner radius of 30 cm, giving a
3 cm clearance between the inner and outer shells on all sides. The
inner shell was assumed to be at a voltage $V$ relative to the outer
shell, which was assumed to be grounded. In an actual experiment,
$V$ will likely be hundreds of kV. But because all of the potentials
we calculated scale linearly with $V$, we set $V=1\textrm{ V}$ for
our calculations, and then scaled the results.

For simplicity we performed our calculations for a cylindrically-symmetric
geometry, replacing the holes in the inner conductor's end caps with
radial slices generated by rotating a 1 cm hole about the axis at
a radius of 25 cm. As such, the fringing fields that we calculate
will be significantly larger than the actual fields in the real apparatus,
and the calculation should be considered a {}``worst case'' estimate.

\section{Equations for Classical and Non-classical Potentials}

To find the classical fringing-field potential $\phi_{c}$ from a
given set of static boundary conditions we use Laplace's equation.
For the non-classical field, we use the counterpart to Laplace's equation
for massive photons: \begin{equation}
\nabla^{2}\phi-\mu^{2}\phi=0.\label{eq:Proca}\end{equation}
 The constant $\mu$ in this equation is related to the photon rest
mass $m_{\gamma}$ by the relation $\mu=m_{\gamma}c/\hbar$.

For the non-classical calculation, instead of solving for $\phi$,
we solved for the deviation from the classical potential $\phi_{p}=\phi-\phi_{c}$.
With this definition, Eq.\ \ref{eq:Proca} becomes\begin{equation}
\nabla^{2}\phi_{c}+\nabla^{2}\phi_{p}-\mu^{2}\phi_{c}-\mu^{2}\phi_{p}=0.\end{equation}
 Laplace's equation states that $\nabla^{2}\phi_{c}=0$, so the first
term in the above equation is zero. Also, since $\mu$ is known to
be very small, we expect $\phi$ to be approximately equal to $\phi_{c}$.
This implies that $\phi_{p}$ will be very small compared to $\phi_{c}$.
As such, we can drop the last term. And since we don't know a-priori
what $\mu$ is equal to, we will normalize our equation by defining
a new parameter $\phi_{s}=\phi_{p}/\mu^{2}$. Making this substitution
and cancelling out the $\mu^{2}$ in both terms we get \begin{equation}
\nabla^{2}\phi_{s}=\phi_{c}\label{eq:phi_s}\end{equation}

Equation \ref{eq:phi_s} is simply Poisson's equation with $\phi_{c}$
playing the part of the charge distribution. As such, our simulation
need only to be able to solve one equation, \begin{equation}
\nabla^{2}\phi_{x}=y.\label{eq:Poisson}\end{equation}
To calculate the classical potential $\phi_{c}$, we simply replace
$\phi_{x}$ with $\phi_{c}$ and insert $y=0$ into this equation.
To calculate the non-classical part of the potential, we first calculate
$\phi_{c}$. Then we then replace $\phi_{x}$ with $\phi_{s}$, and
insert our previously calculated values for $\phi_{c}$ as the source
term $y$. For an axially-symmetric system, Eq.\ \ref{eq:Poisson}
can be written as a two-dimensional equation in cylindrical coordinates:\begin{equation}
\frac{1}{r}\frac{\partial\phi_{x}}{\partial r}+\frac{\partial^{2}\phi_{x}}{\partial r^{2}}+\frac{\partial^{2}\phi_{x}}{\partial z^{2}}=y,\label{eq:PoissonAxial}\end{equation}
where $r$ and $z$ are the radial and axial coordinates.

Note that if $\mu\neq0$, a constant potential is not a solution to
Eq.\ \ref{eq:Proca}, and we are not free to arbitrarily define the
outer conductor to be at $V_{g}=0$. But if this conductor is Earth
grounded, due to the huge capacitance of the Earth it is reasonable
to assume that $V_{g}$ will remain fairly constant as the voltage
on the inner conductor is changed. In the proposed experiment the
inner conductor's voltage relative to the outer conductor would be
periodically reversed. Rather than measuring the field inside the
conductor, the \emph{difference} between the field before and after
the reversal would be measured. In this measurement the the unknown
voltage offset $V_{g}$ will cancel. As such, setting $V_{g}=0$ in
our calculations still gives us meaningful results.

\section{Methodology}

We did our calculations using a finite difference method, in which
the potential is calculated at points on a grid. Derivatives are approximated
to second order from adjacent points on the grid. Our grid was evenly
spaced in both dimensions, with points separated by a distance $a$.
The potential $\phi_{x}$ at each point on the grid will be written
as $\phi_{i,j}$ and the source term $y$ at each point as $y_{i,j}$,
where $i$ and $j$ are integers labeling the point. We will define
$i=j=0$ at the center of the conducting shells, such that the actual
coordinates of each grid point are equal to $r_{i}=ai$ and $z_{j}=aj$.
Using these definitions, Eq.\ \ref{eq:PoissonAxial} can be approximated
by the equation\begin{eqnarray}
\frac{\phi_{i+1,j}-\phi_{i-1,j}}{2ar_{i}}+\frac{\phi_{i+1,j}-2\phi_{i,j}+\phi_{i-1,j}}{a^{2}}\nonumber \\
+\frac{\phi_{i,j+1}-2\phi_{i,j}+\phi_{i,j-1}}{a^{2}} & = & y_{i,j}.\label{eq:PoissonFinite}\end{eqnarray}
This can be solved for $\phi_{i,j}$ in terms of the known quantity
$y_{i,j}$ and the value of $\phi_{x}$ at adjacent grid points:\begin{eqnarray}
\phi_{i,j} & = & \frac{1}{4}\bigg[\left(1+\frac{a}{2r_{i}}\right)\phi_{i+1,j}+\left(1-\frac{a}{2r_{i}}\right)\phi_{i-1,j}\nonumber \\
 &  & \,\,\,\,\,\,\phi_{i,j+1}+\phi_{i,j-1}-a^{2}y_{i,j}\bigg].\label{eq:phiij}\end{eqnarray}

We began our simulation with an arbitrary value of $\phi_{i,j}$ at
each grid point. Then using Gauss-Seidel iteration \cite{NumericalRecipesInC},
this equation was evaluated at each point to produce an updated value
of $\phi_{i,j}$. After many iterations, $\phi_{i,j}$ eventually
converged to the correct values to solve Eq.\ \ref{eq:PoissonFinite}.
To accelerate convergence, we used the successive over-relaxation
method \cite{NumericalRecipesInC}.

One group of points which requires attention are the points for which
$i=0$. Because $r$ is always positive in cylindrical coordinates,
$\phi_{i,j}$ is not defined for negative values of $i$. But there
is effectively no difference between cylindrical and Cartesian coordinates
for the row of points along the axis. So for these points we calculated
derivatives using Cartesian coordinates knowing that the points directly
below the axis and just into or out of the two-dimensional grid should
have the same value as the point directly above each point on the
axis. This gives the equation\begin{equation}
\phi_{0,j}=\frac{1}{6}\left(4\phi_{1,j}+\phi_{0,j+1}+\phi_{0,j-1}\right).\label{eq:phi0j}\end{equation}

In addition to axial symmetry, the conductors have mirrored symmetry
about their center. This allows us to throw away all of the grid points
with $j<0$, cutting the number of grid points in half. Doing this
requires us to treat the $j=0$ points differently, because our grid
no longer contain values for $\phi_{i,-1}$. By symmetry we know that
$\phi_{i,j}=\phi_{i,-j}$. This allows us to replace $\phi_{i,-1}$
with $\phi_{i,1}$ in Eq.\ \ref{eq:phiij}. The point $\phi_{0,0}$
is a special point, being both a member of the $r=0$ and $z=0$ groups
of points. For this point we use Eq.\ \ref{eq:phi0j}, but substitute
$\phi_{0,1}$ for $\phi_{0,-1}$.

After every 100 iterations, the program calculated an estimated error
at each point by evaluating Eq.\ \ref{eq:phiij} at each point without
changing any values on the grid. We defined the estimated fractional
error to be\[
\left|\frac{\phi_{i,j}-\Phi_{i,j}}{\phi_{i,j}}\right|,\]
where $\phi_{i,j}$ is the actual value at grid point ($i$,$j$)
and $\Phi_{i,j}$ is the value calculated from Eq.\ \ref{eq:phiij}.
If there were no points on the grid with a fractional error larger
than $10^{-8}$, the simulation terminated.

\begin{figure}
\begin{center}\includegraphics[%
  width=3cm,
  keepaspectratio,
  angle=-90]{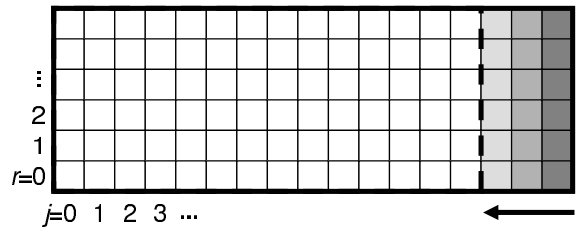}\end{center}

\caption{Reducing the grid size. In this process the fractional error of the
outermost column of points (the dark grey column) is evaluated. If
the fractional error for each of these points is less than a set value,
the grid size is reduced and the next column of points (the medium
grey column) is evaluated. If the error for each point in this column
is smaller than the set value, the grid is reduced again and the next
column (the light grey column) is evaluated. This process continues
until a column is found which has at least one point with a fractional
error larger than the specified limit. In this example, the right-most
white column was the first column found to have a point with a fractional
error above the set limit, and the grid was reduced in the direction
of the arrow to the dashed line. \label{cap:Reducing-the-grid}}
\end{figure}

While experimenting with different conductor geometries, we significantly
decreased the required computation time by successively reducing the
size of the grid in the axial direction. This was done by finding
columns which had already converged near their final value and then
using these values as the new boundary conditions for a problem involving
a smaller grid. To do this the program used a variable $jmax$, initialized
to the largest $j$ index on the grid, and only updated points with
$j\leq jmax$. After each set of 100 iterations the program stopped
and calculated the fractional error for each of the points with $j=jmax$.
If the fractional error at every point in this column was smaller
than $10^{-8}$, the program would reduce $jmax$ by one, thereby
reducing the effective size of the grid. The program then repeated
this process until it found a column which contained at least one
point who's fractional error was larger than the specified value.
Then the over-relaxation parameter was re-calculated and the next
set of 100 iterations was performed. This process is illustrated in
Fig.\ \ref{cap:Reducing-the-grid}.

The error introduced by stepping in should be negligible; if a column
has converged to within a factor $f$ of its final value, the error
introduced onto other points by {}``freezing'' this column should
be of order $f$. Therefore, for a grid with $N_{z}$ columns, the
largest fractional error introduced anywhere on the grid by this method
should be on the order of $f\sqrt{N_{z}}$ if the errors are assumed
to be random, and on the order of $fN_{z}$ or smaller otherwise.

Because we have not done a rigorous theoretical study of this method,
once we had decided on the final geometry for our conductors we verified
our calculations by performing additional computations which did not
use this {}``stepping in'' method. The result of these calculations
were identical to those done with the stepping-in technique to the
seven digits of precision saved at the end of the calculations.

\section{Classical Fringing-Field Potential}

For the classical fringing-field calculation we are mainly interested
in how the potential inside the inner conductor varies from $V$,
the voltage of the inner conductor. To keep round-off error from completely
masking these variations, we made use of the fact that classical electromagnetism
allows us to arbitrarily add a constant potential. So rather than
solving Laplace's equation for $\phi_{c}$, we instead solved for
$\Delta\phi_{c}=\phi_{c}-V$ using the same equation but different
boundary conditions --- $-V$ on the outer conductor and $0$ on the
inner conductor. This way we found small deviations from zero rather
than a finite potential.

\begin{figure}
\begin{center}\includegraphics[%
  clip]{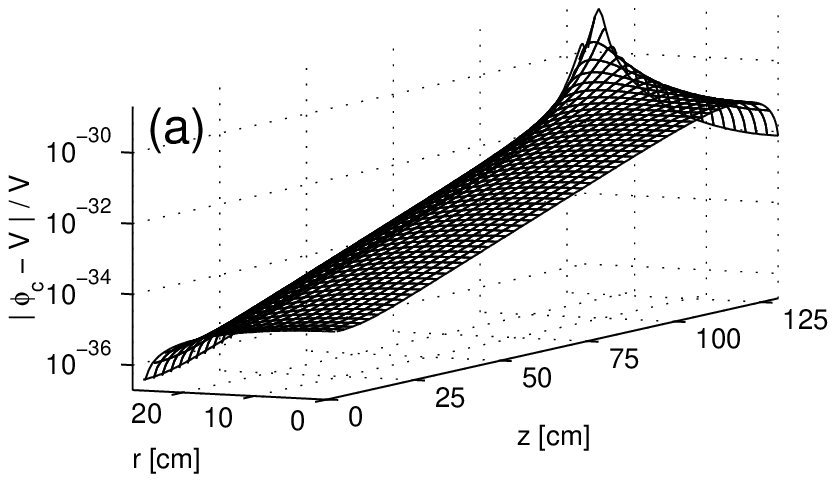}\end{center}

\begin{center}\includegraphics[%
  clip]{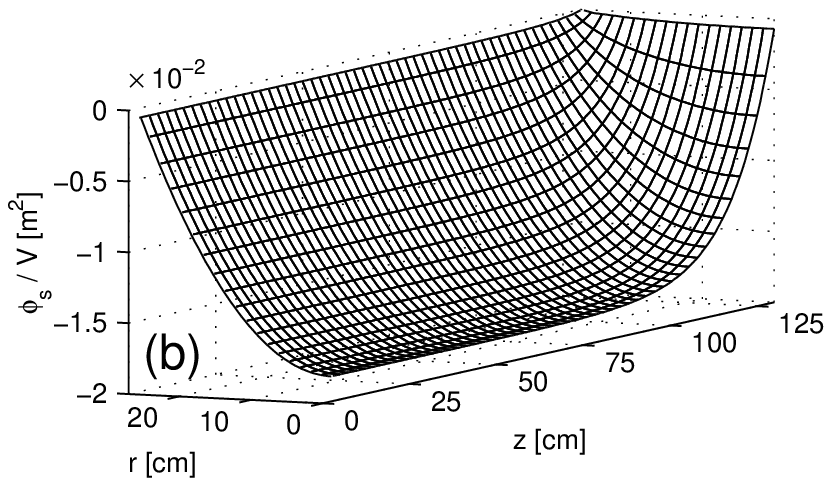}\end{center}

\begin{center}\includegraphics[%
  clip]{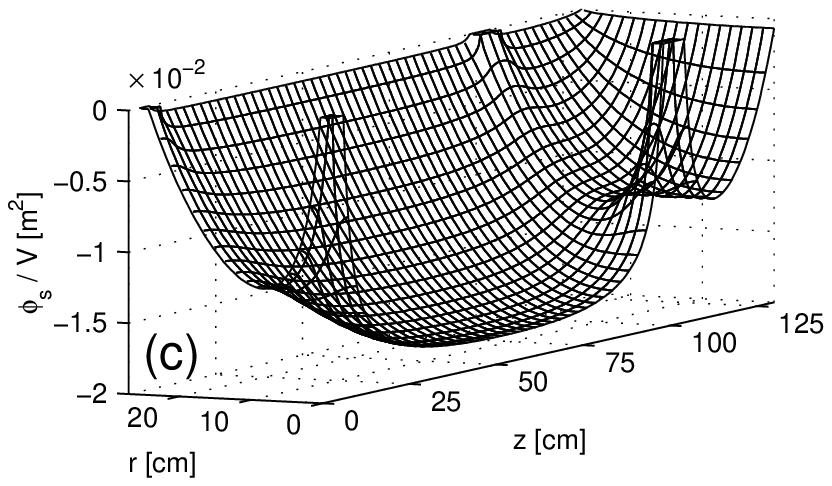}\end{center}

\caption{The results of the numerical calculations. All calculations were
done with a grid spacing of $a=1.67\times10^{-2}$ cm. Only the region
inside the inner conductor is shown. In (a) a semi-log plot of the
classical fringing-field potential is shown. The peak in the upper
right-hand corner is at the location of the radial slice in the end
cap. In (b) the non-classical potential $\phi_{s}$ is plotted. Frame
(c) shows the effect of additional conducting objects inside the inner
shell. \label{cap:fields}}
\end{figure}

The results of this calculation are shown in Fig.\ \ref{cap:fields}(a).
We have verified that the errors due to a finite grid are negligible
by performing this calculation with other grid spacings. If the grid
spacing $a$ is increased by a factor of 3, $\Delta\phi_{c}$ at the
center of the conductors changes by 9\%. If $a$ is increased by a
factor of 1.5, it changes by only 1.3\%. We fit these three results
to a power law, and found that the fit is in good agreement with a
fourth data set data in which $a$ is increased by a factor of 6.
From this fit we estimate an error on the order of 0.3\%.

To verify the validity of our results we used a series solution to
calculate the field inside of the inner conductor assuming that the
inner conductor was grounded, and that the potential inside of the
radial slices in the end caps was a fixed constant. This calculation
showed that the potential at the center of the tube would be $4.5\times10^{-7}$
times the potential inside the radial slices, in good agreement with
our numerical calculation. We also found a series solution for the
field between two grounded cylinders with a fixed potential applied
to the gap at the ends. The length of the cylinders was assumed to
be equal to the sum of the lengths of the two end caps, with a gap
between them equal in size to the radial slice in the end caps. The
field in the center of the gap turned out to be $2\times10^{27}$
times smaller than the potential at the end, also in agreement with
our numerical calculation.

\section{Non-Classical Potential}

To calculate $\phi_{s}$, we set $y=\phi_{c}$ in Eq.\ \ref{eq:phi_s}
and used the same methods discussed above. Since the potential on
the conducting surfaces is given, the deviation from the classical
potential should be zero on these surfaces. So the boundary conditions
for the $\phi_{s}$ calculation were that $\phi_{s}=0$ on both of
the conductors.

The source term $y$ in this calculation is equal to $\phi_{c}$.
We can obtain this by adding $V$ to our already completed calculation
of $\Delta\phi_{c}$. Inside the inner conductor, where $\Delta\phi_{c}$
is very small, this results in values accurate to the total precision
allowed by the double-precision floating-point format. But between
the two conductors, adding $V$ to $\Delta\phi_{c}$ results in \emph{lost}
precision. We are only interested in the fields inside the inner conductor,
which should not be affected by this lost precision --- the non-classical
potential is predominantly generated by the local source term $y$
rather than from fringing fields generated outside the inner conductor.
But to be extra careful we also calculated $\phi_{c}$ directly using
the correct boundary conditions (the inner conductor at $V$ and the
outer one at $0$), and used these values wherever $\phi_{c}$ was
less than $V/2$.

The results of this calculation are shown in Fig.\ \ref{cap:fields}(b).
Figure \ref{cap:fields}(c) shows the results of a similar calculation
in which additional conductors were added to simulate optics and other
objects inside the inner shell. These conductors consisted of a ring
and a cylinder at the three locations where gratings would be positioned
in the interferometer. The rings were assumed to be 6 cm wide and
1.5 cm thick, just touching the inner shell, and the cylinders to
be 6 cm long and 1.5 cm in radius, centered on the axis of the inner
shell, as shown in Fig.\ \ref{cap:Layout-of-conductors.}. These
conductors were assumed to be at the same potential as the inner shell.

We estimate the error in these calculation to be very small --- to
the seven digits output by our program there was no change in $\phi_{s}$
when we increased the grid spacing by a factor of 3.

\section{Implications}

\begin{figure}
\begin{center}\includegraphics[%
  clip]{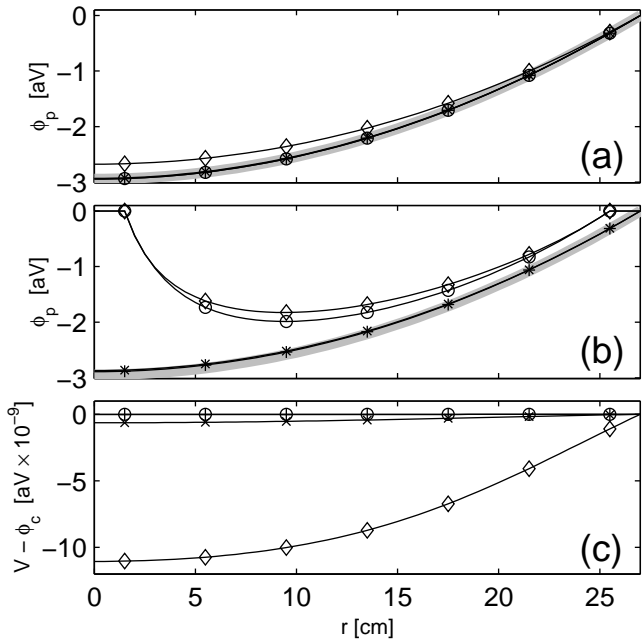}\end{center}

\caption{Comparing the different potentials. The potential is plotted as a
function of radius at four different points along the axis. In (a)
the non-classical field is shown. In (b) the non-classical field is
shown for the case in which additional conductors have been placed
inside the inner shell. In (c) the classical field is shown. The lines
marked with circles and diamonds are at axial distances of 0 and 100
cm from the center of the conductors --- the locations of the gratings.
The lines marked with $+$ and $\times$ are at axial distances of
33 and 67 cm, respectively. All potentials were plotted assuming that
$V=200\,\textrm{kV}$ and $\mu=2.8\times10^{-11}\textrm{ m}^{-1}$.
The thick grey lines in (a) and (b) represent the analytical solution
for the non-classical field inside of an infinite cylinder. \label{cap:Comparing-Potentials}}
\end{figure}

The fields are compared in Fig.\ \ref{cap:Comparing-Potentials}.
While this figure shows the \emph{potential} at \emph{each point}
moving radially from the axis, for the proposed ion interferometer
experiment, all that is important is the \emph{field} \emph{at} \emph{the
location of the ion beam}. As such, the most important information
in this figure is the radial slope of the potential at $r=25\textrm{ cm}$.

In Fig.\ \ref{cap:Comparing-Potentials}(a) we see that the non-classical
field is approximated very well by the field of an infinite cylinder.
In Fig.\ \ref{cap:Comparing-Potentials}(b) we see that although
the effect of small objects inside the shell on the non-classical
potential is not negligible, it should not greatly change the sensitivity
of the experiment as long as care is taken. Because axial symmetry
is assumed, the additional conductors took the form of large rings
rather than small rectangles which would better approximate an optical
mount, and this plot can be considered an extreme {}``worst-case''
estimate.

Note that the vertical axis in Fig.\ \ref{cap:Comparing-Potentials}(c)
is about $2.4\times10^{8}$ times smaller than in (a) and (b), indicating
that the fringing-fields inside the inner conductor should be completely
negligible for values of $\mu$ much smaller than $2.8\times10^{-11}\textrm{ m}^{-1}$(corresponding
to a photon rest mass of $1\times10^{-53}\textrm{ kg}$, over 600
times smaller than the current experimental limit measured in \cite{Crandall83}).
Consequently, fringing fields should not be a problem in the proposed
experiment.

\section{Conclusions}

In conclusion, we have conducted a numerical study of classical and
non-classical electrostatic potentials in an axially-symmetric nested
conductor configuration. The results show that the assumptions in
our recently proposed ion-interferometry experiment are valid. The
calculations show that for values of $\mu$ much smaller than the
current experimental limit, non-classical fields should still dominate
over fringing fields. Furthermore, we have shown that the non-classical
field approximates the simple field of an infinitely long set of conductors.

We would like to acknowledge Ross Spencer for his help on every aspect
of this study. This work was funded by BYU's Office of Research and
Creative Activities.

%\bibliographystyle{L:/Research/Publications/PhotonMassCalcs/apsrev}
%\bibliography{L:/Research/Publications/PhotonMassCalcs/PhotonMassBib14}

\end{document}